# Using micro- and macro-level network metrics unveils top communicative gene modules in psoriasis

Reyhaneh Naderi [1, †], Homa Saadati Mollaei [2], Arne Elofsson[3] and Saman Hosseini Ashtiani [3, †, *]

[1] Department of Artificial Intelligence and Robotics, Faculty of Computer Engineering, Iran University of Science and Technology, Tehran, Iran.; r_naderi@alumni.iust.ac.ir
[2] Department of Advanced Sciences and Technology, Islamic Azad University, Medical Branch, Tehran Iran.; h.saadati1995@gmail.com
[3] Department of Biochemistry and Biophysics and Science for Life Laboratory, Stockholm University, 106 91 Stockholm Sweden; arne.elofsson@dbb.su.se
* Correspondence: saman.hosseini-ashtiani@dbb.su.se; Tel.: +46-762623644
† These authors contributed equally to this work and should be considered joint first authors.

**Abstract**

**Background:** Psoriasis is a multifactorial chronic inflammatory disorder of the skin with significant morbidity, characterized by hyper proliferation of the epidermis. Even though psoriasis etiology is not fully understood, it is believed to be multifactorial with numerous key components. **Methods**: In order to cast light on the complex molecular interactions in psoriasis vulgaris at both protein-protein interactions and transcriptomics levels, we analyzed a set of microarray gene expression analysis consisting of 170 paired lesional and non-lesional samples. Afterwards, a network analysis was conducted on protein-protein interaction network of differentially expressed genes based on micro- and macro-level network metrics at a systemic level standpoint. **Results**: We found 17 top communicative genes, all of which experimentally proven to be pivotal in psoriasis were identified in two modules, namely, cell cycle and immune system. Intra- and inter-gene interaction subnetworks from the top communicative genes might provide further insight into the corresponding characteristic mechanisms. **Conclusions**: Potential gene combinations for therapeutic/diagnostics purposes were identified. Moreover, our proposed pipeline could be of interest to a broader range of biological network analysis studies.

## 1. Introduction

Psoriasis is a multifactorial chronic inflammatory disorder of the skin, affecting about 2–3% of the population worldwide characterized by hyper proliferation of the epidermis. The most prevalent subtype of the disease is psoriasis vulgaris (plaque psoriasis). Even though psoriasis etiologies are not fully understood, it is believed to be multifactorial with numerous key components including genetic susceptibility, environmental triggers in combination with skin barrier disruption and immune dysfunction. Regarding the pathogenesis and the molecular mechanisms responsible for psoriasis, many details are still unclear. There are a host of treatments for moderate to severe psoriasis patients ranging from topical creams to systematic drugs and phototherapy. However, our inability to thoroughly understand the disease prevents the optimal use of these therapeutic routes and there is no satisfactory treatment for most psoriatic patients yet [1, 2]. Generally, disease genes have a tendency to show tissue-specific expression patterns, be over-expressed and have an increased mutation rate with respect to evolutionary time [3]. Several computational studies have tried to find and elucidate potential contributing genes in psoriasis. Network analysis can be used to mine interconnections between large amounts of data from biological and medical science. Such methods can give rise to new hypotheses and potential knowledge that would not be easily decrypted otherwise. The characterization of biological networks by means of network topological properties is widely used for gaining insight into the global and local interaction properties [4]. Li et al. performed weighted gene co-expression network analysis (WGCNA) on RNA-Seq data of psoriatic and normal skin tissues to detect the key long non-coding RNAs (lncRNAs) and mRNAs associated with psoriasis. They recognized important lncRNAs in lncRNA-mRNA co-expression network based on the degree distribution of the network [5]. Zhang et al. analyzed microarray data from plaque psoriasis samples revealing that metabolic and viral infection-associated pathways were enriched the most. The hub genes, which were identified to be implicated in playing key regulatory roles in the Protein-Protein Interaction (PPI) network, were determined based on the overlapping outcomes achieved by Maximal Clique Centrality (MCC) and Density of Maximum Neighborhood Component (DMNC) topological analysis methods [6]. Mei et al. used microarray data in psoriatic patients to construct co-expression modules based on the differential expression of various disease-characteristic genes, and screened the likely effective therapeutic molecules and their possible target. Furthermore, the degree centrality measure was used to identify the network hub genes [7].

Sundarrajan et al. conducted an integrated systems biology approach to understand the molecular alliance of psoriasis with its corresponding comorbidities. A microarray dataset was fetched for each disease for this study. "Network medicine" was applied to scrutinize the molecular intricacy of psoriasis leading to the identification of new molecular associations among apparently distinct clinical manifestations [8]. Arga et al. used topological, modular and a novel correlation analysis based on fold changes of the microarray data fetched from three different platforms. Psoriasis associated PPI networks was based on a dual-metric approach using degree and betweenness. They highlighted the significance of JAK/STAT pathway in psoriasis along with proposing potential psoriatic biomarkers [9]. Piruzian et al. applied a number of network and meta-analysis techniques on paired lesional and healthy microarray datasets to reveal the similarities and dissimilarities between proteomics- and transcriptomics-level perturbations in psoriasis. They particularly tried to reveal novel regulatory pathways in psoriasis development and progression [10].

In this study, we performed a network topological investigation of highly communicative genes and mechanisms underlying medium-severe psoriasis. We looked into this disease at a systemic level from a network-based perspective to characterize the patterns of mRNA gene expression data in the context of network and subnetwork analysis results. To the best of our knowledge this is the first study to consider as many micro- and macro-level network metrics as possible in order to investigate the most important long- and short-range interactions of the PPI network in moderate to severe psoriasis vulgaris.

**2. Materials and Methods**

A complete workflow of our study design is shown in Figure 1.

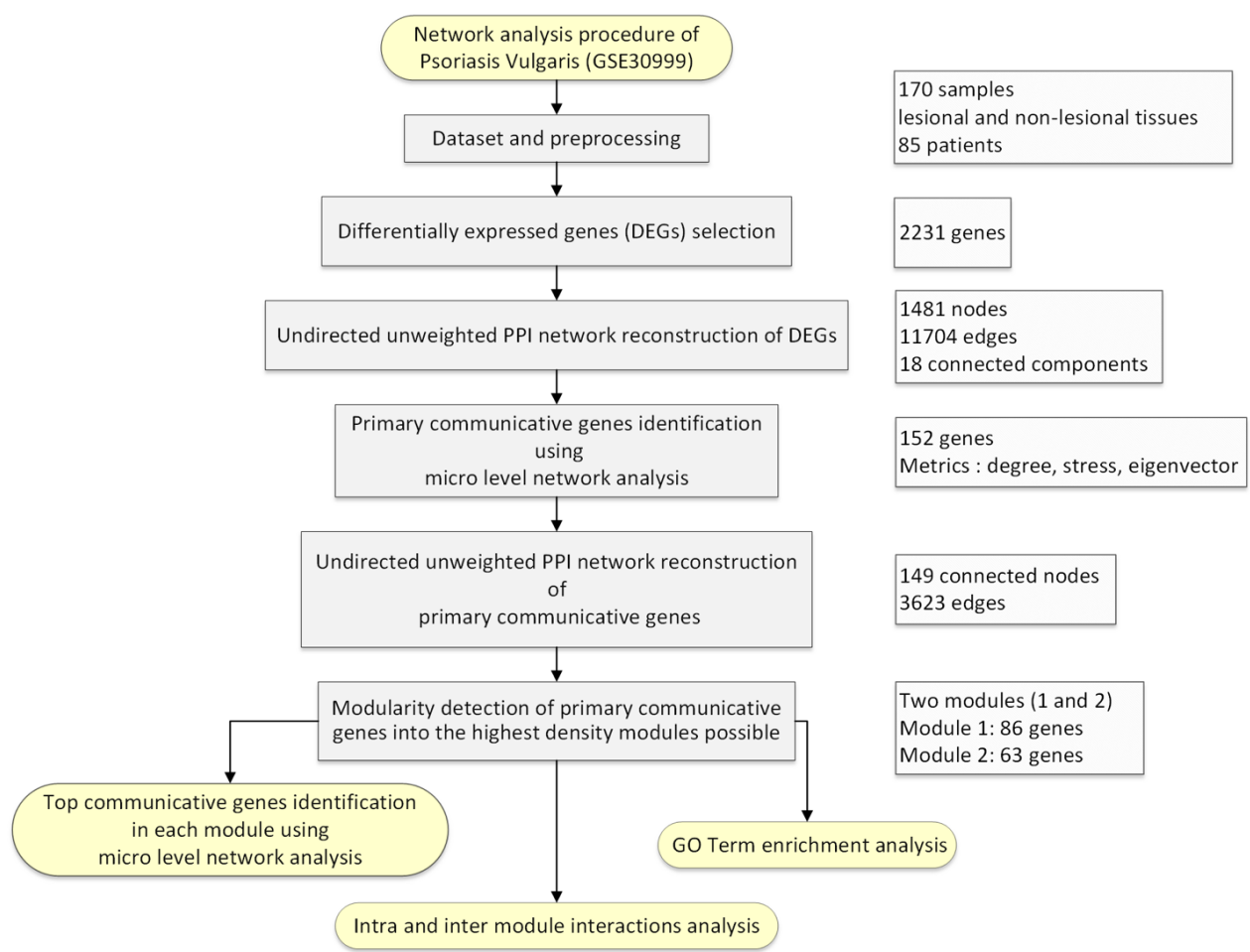

**Figure 1**. The workflow of study design of network analysis in psoriasis vulgaris patients' dataset (GSE30999) downloaded from the Gene Expression Omnibus (GEO).

2.1. Dataset

The microarray dataset used in this paper was generated by Suárez-Farinas et al. [11] based on the GPL570 Affymetrix Human Genome U133 Plus 2.0 array platform and is available in NCBI's Gene Expression Omnibus (GEO) repository, GSE30999 (https://www.ncbi.nlm.nih.gov/geo/query/acc.cgi?acc=GSE30999). It was selected from 85 patients with moderate to severe psoriatic patients. The patients had not received any active psoriasis therapy. Considering the large enough number of 170 psoriasis vulgaris paired samples in this dataset (greater than 30 according to the central limit theorem) in comparison with other available paired tissue datasets, the mentioned dataset was chosen. The use of a much larger sample dataset facilitates a more comprehensive perspective of the processes affecting the psoriatic expression profile. Moreover, we did quality control (QC) on the dataset and the result was satisfactory (Figure 2-4). We eschewed combining different array experiments data from the NCBI's GEO series in order to avoid the variations associated with batch effects from merging different datasets of multiple microarray experiments and involving different population groups, particularly because psoriasis is a multifactorial disease with a combination of key environmental and genetic factors [1, 12].

2.2. Preprocessing

In this step, we performed some QC (Figure 2-4) checks involving sample-level plots to ensure that the samples are statistically in good conditions. These checks comprise Principal Component Analysis (PCA) plot, density function plot, heatmap of Pearson's correlation coefficient (r2) between samples and boxplot. To assess overall similarity and sources of variations between the samples, we performed PCA technique and Pearson's correlation values heatmap plot. All pairwise Pearson correlations on samples were calculated by QC tool accessible from Expression Console software, version 1.4 (Affymetrix). For a sample size > 30 (here 170) according to the central limit theorem, we assumed a population in which samples have normal distribution for a hypothesis testing. PCA transforms gene expression level in principal component space, reducing each sample to a single point. It helps us distinguish samples using expression variations and determine whether the lesional and non-lesional samples are differentiable after normalization (Figure 4). To stabilize variances and make the gene expression levels less dependent on the absolute values, the gene expression levels were transformed to logarithmic base 2 scale. Because of utilizing a linear model in our statistical procedure, which depends on the assumption of normality, the dataset was normalized. We applied MAS 5.0 algorithm [13] implemented by Affymetrix available in affy package in Bioconductor. The preprocessing was performed using R (version 3.6.1).

2.3. Selection of differentially expressed genes (DEGs) in psoriasis vulgaris patients

The samples are defined as non-lesional (NL) and lesional (LL) skin. To fit the linear model to each gene given in a series of arrays in the samples, a design matrix was created, a contrast matrix was then computed from the linear model fit. Finally, the differentially expressed genes between NL and LL samples were assessed using a moderated t-test called empirical Bayesian method [14]. The Proportion parameter of genes for eBayes algorithm was considered 0.01. The cutoff criteria of p-value <0.01 and |logFC| ≥1 were considered as the thresholds for the significance to extract DEGs among 54675 probe sets. The p-values were adjusted using the default adjustment method named BH method described by Benjamini and Hochberg (false discovery rate (FDR)) [15]. The mentioned method is very likely to be appropriate for microarray studies. Thus, the criterion of adjusted p-value<0.01 was also applied. Therefore, 2231 unique genes were identified as DEGs. Afterwards unique genes were ranked with respect to B column (log-odds that the gene is differentially expressed) and selected top 2000 genes (Table SI) in order to submit to STRING [16] database. The hgu133plus2.db annotation package [17] was used to transform probe IDs into gene symbols. For one gene symbol corresponding to several probe IDs, the maximum absolute value of logFC for probes was measured as the final value. All the procedures up to this point was processed under the R statistical environment (version 3.6.1, https://www.r-project.org) with the use of the package limma in Bioconductor release 3.9.

2.4. PPI undirected unweighted network reconstruction of DEGs

To reconstruct the network of PPI data, we used STRING database (a search tool for the retrieval of PPIs) version 10.5 (https://string-db.org) covering more than 2000 organisms including direct (physical) as well as indirect (functional) associations. Thus, DEGs were submitted to STRING with minimum required interaction score 0.4 and confident score ≥ 0.15. The proteins were represented as nodes in the network, whereas, their degrees corresponded to the number of edges associated with that node [16].

2.5. Primary communicative genes identification in PPI network

We topologically analyzed psoriasis associated PPI network imported from STRING as undirected and unweighted. 1481 nodes and 11704 edges were detected using NetworkAnalyzer [18], which is a Java plugin integrated into network software tool Cytoscape 3.7.2 [19]. This popular tool quantifies the specific parameters describing the network topology to identify primary communicative genes. In this study, the influence of the genes in the network was characterized and quantified based on their position in the network which was described mathematically using centrality measures, including local scale (degree) and global scales (stress, betweenness, closeness and eigenvector). We considered them as micro-level network metrics. In network science, the mentioned metrics are directly proportional to the topological importance of the nodes [20]. More than 10 percent of the total number of 1481 nodes, being derived with respect to the mentioned metrics, were called primary communicative genes (Figure 4a). As a result, a minimum threshold of degree 50, stress 200,000 and eigenvector 0.05 were considered. By the union of three gene sets, 152 genes called primary communicative genes were obtained. Eigenvector metric was calculated by CentiScaPe plugin [21].

2.6. Undirected unweighted PPI network reconstruction and analysis of primary communicative genes

As described in previous section, we extracted the primary communicative genes by micro-level network metrics including some local and global centrality measures based on undirected unweighted network analysis. These genes were identified according to the interactions between the nodes associated with the 1445-node connected network. We hypothesized that these genes could be a subset of influential nodes within the mentioned network, hence, the dependencies of these genes to themselves could be more likely to be important than their dependencies to other nodes in the 1445-node network. Therefore, we merely considered the relationships between these genes themselves. We mapped primary communicative genes to STRING database to obtain interactive relationships among 152 genes. The primary communicative genes PPI network consisted of 149 connected nodes with average node degree 47, three single nodes (DNAH8, NALCN, PNPLA7) and 3623 edges. The 149-node network was analyzed with the metrics, namely, degree, stress, betweenness, eigenvector and closeness using network software tool Cytoscape 3.7.2 (Figure 5).

2.7. Macro level network analysis followed by top communicative genes identification

Modularity is one of the topological features of interconnected nodes evaluating the quality of the modules resulting from modularity detection techniques in the context of network science. The modularity detection problem considers decomposing a network into modules of densely connected nodes [22]. To further topologically characterize the PPI network of primary communicative genes, we used a modularity detection algorithm based on a heuristic method proposed in [22] that uses modularity optimization. We tried to maximize modularity to gain the highest density modules possible (Figure 4b). The micro level network metrics obtained from the analysis of the primary communicative genes PPI network, mentioned in previous section, were then applied for each module to identify top communicative genes. Further investigations on gene sets in each module in terms of metrics showed that some metrics (closeness in modules 1 and 2, eigenvector in module 1) had very low variance in their corresponding values for the nodes in such a way that those metrics could not be used as discriminating criteria. Consequently, the top communicative genes from each module were extracted with the parameters of degree, stress, betweenness and eigenvector. To further investigate relationships between top communicative genes themselves in each module and between two modules, we extracted the nested subnetworks involved in top communicative genes in each module (intra module interactions) and between modules (inter module interactions) (Figure 4c,d). We called the modularity detection step with intra and inter module interaction analysis "macro level network analysis". Pathway enrichment analysis was performed by web-based tool Enrichr (http://amp.pharm.mssm.edu/Enrichr/) via WikiPathways [23] and Kyoto Encyclopedia of Genes and Genomes (KEGG) 2019 Human databases [24]. We used Gephi (version 0.9.2), an open-source and multiplatform software [25], to apply the modularity detection procedure and Cytoscape to identify and visualize intra and inter module interactions.

3. Results

3.1. Preprocessing of samples

The Pearson's correlation coefficient (r2) values are presented as a heatmap (Figure S1). Overall, the resulting heatmap indicates a strongly positive correlation between all samples. To obtain microarray data with less bias, data normalization was performed. Figure S2 illustrates less samples medians fluctuations after normalization. PCA plot represents two distinct groups in the dataset after normalization. Accordingly, the samples were separated along the first and the second principal components (PC1, PC2) (Figure 2).

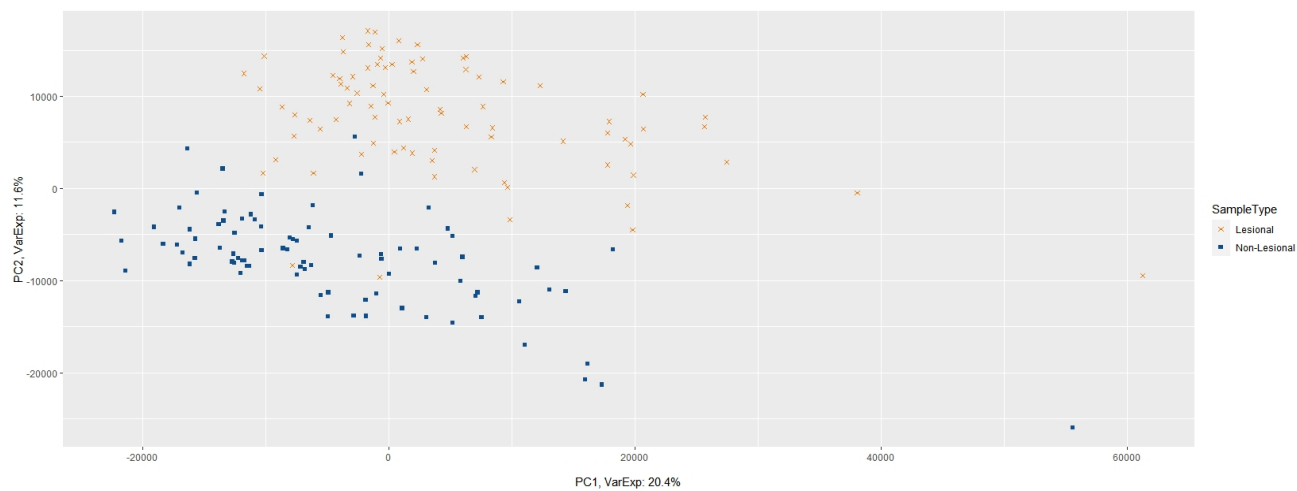

**Figure 2.** The PCA on gene expression values of all samples after normalization. Sample types are highlighted in different colors. PCA analysis shows that the two distinct groups correspond to sample types (lesional or non-lesional) conveying an overall certain discrimination associated with gene expression levels between the two groups. PC1 and PC2 capture 20.4% and 11.6% of the variance in the normalized data. PC1 and PC2 together only reveal 32% of the total variance meaning that other components are also important.

3.2. DEGs in Psoriasis vulgaris patients

2231 DEGs contained 1031 upregulated and 1200 downregulated genes. The top 50 up- or down-regulated genes in psoriasis-associated DEGs are listed in Table S2 and Table S3, respectively. To visualize DEGs in terms of significance and the magnitudes of changes in their gene expression levels as a united plot, we generated a volcano plot (Figure S3).

3.3. PPI network of DEGs reconstruction and analysis

The PPI network in the string database demonstrated 1836 nodes (1481 connected nodes) and 11704 interactions including average node degree 4.8. The network associated with 1481 identified genes was analyzed and reconstructed using Cytoscape (Figure 3). This network contains 18 connected components characterized by one 1445-gene component (network 1), 15 2-gene components (networks 4 to 18) and two 3-gene components (networks 2 and 3). The network statistics has been demonstrated in Figure 3.

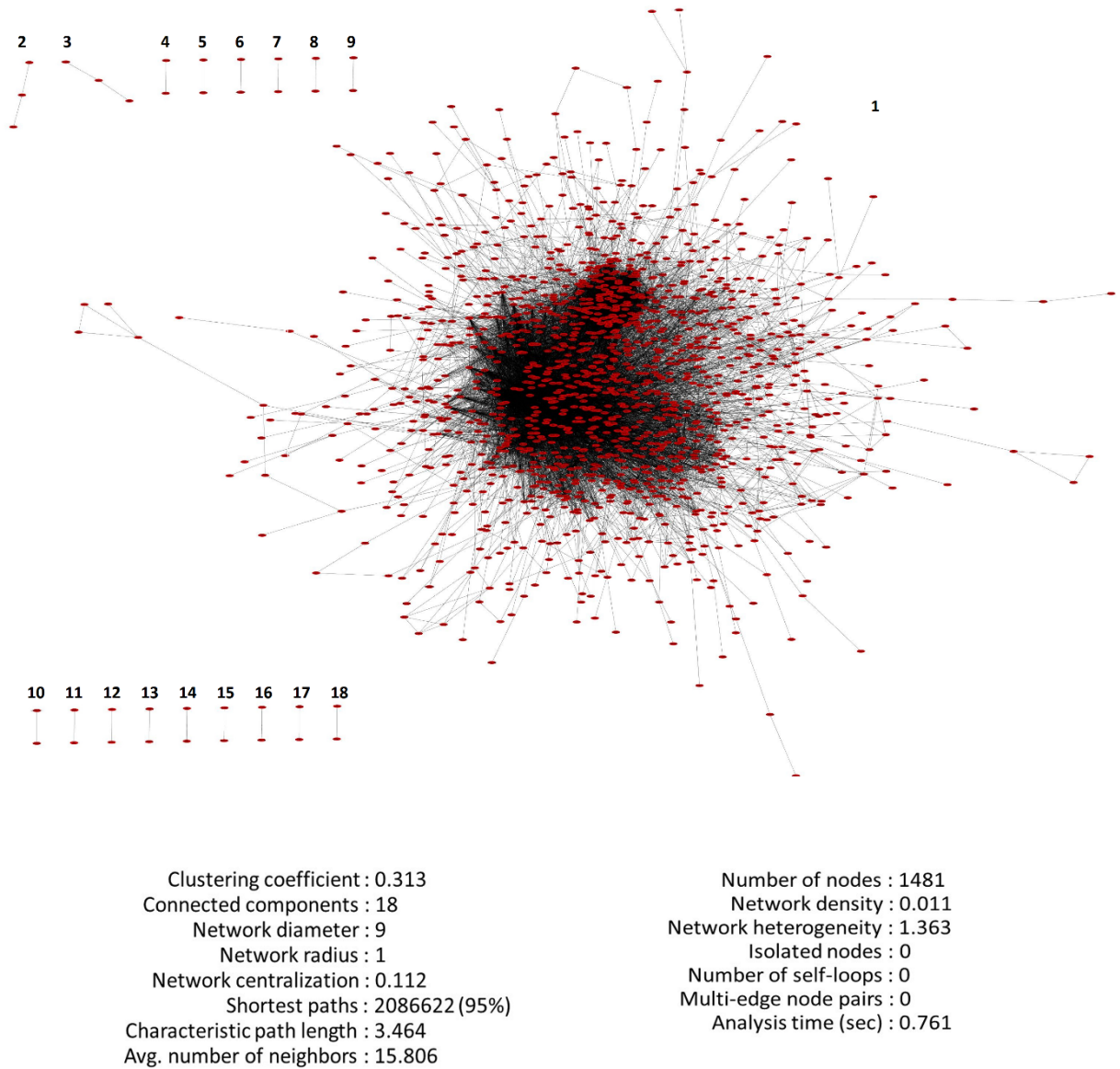

Figure 3. The statistics of 1481-gene PPI network after analysis by Cytoscape.

By analyzing micro level network metrics on main component (component 1 in Figure 3), a total of 152 primary communicative genes were identified among 1481 genes according to combining degree, stress and eigenvector metrics (Figure 4a).

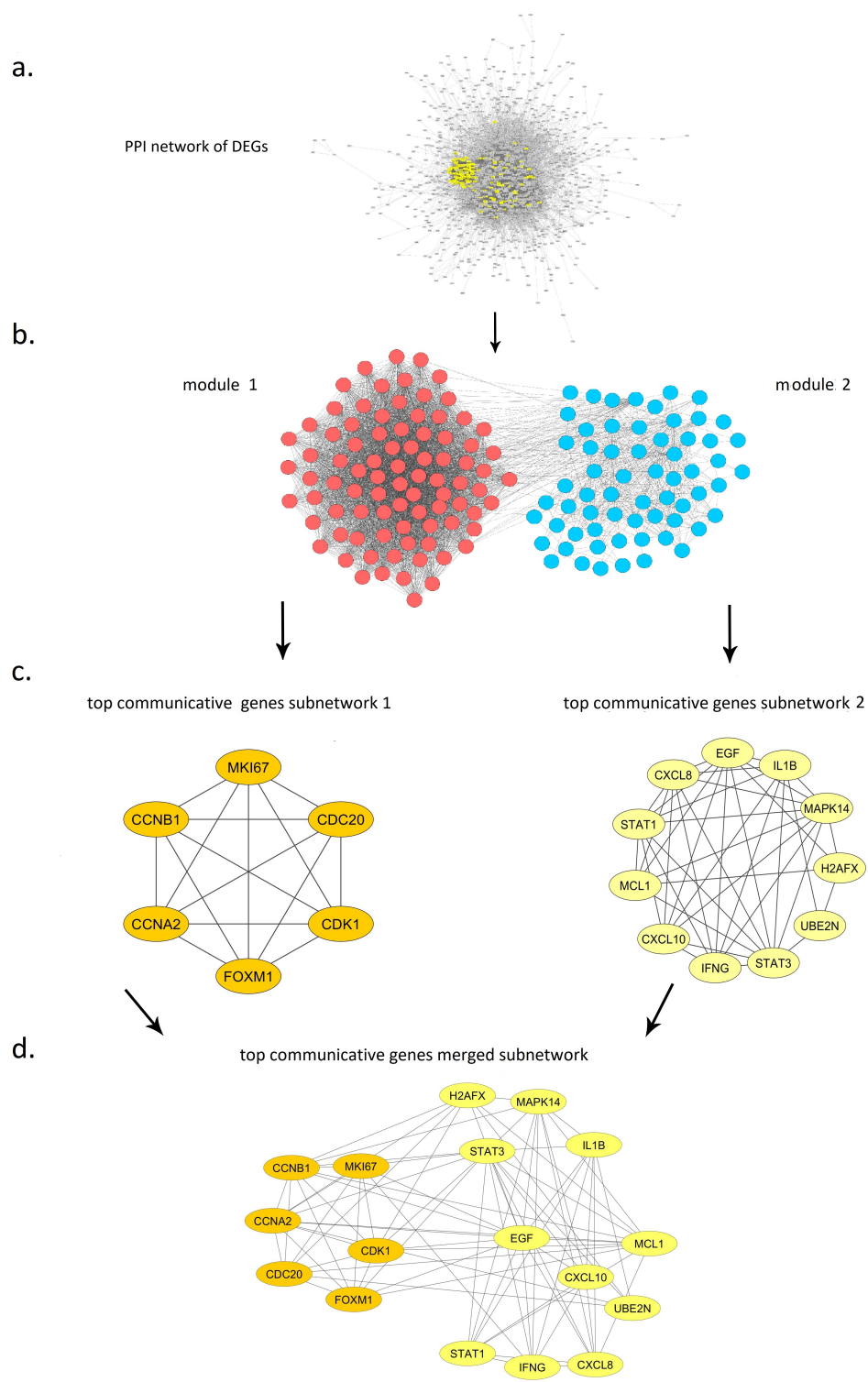

**Figure 4.** Network analysis results. (a) Main component of PPI network of DEGs among 18 components. Selected yellow nodes indicate primary communicative genes based on micro level network analysis. (b) Modularity detection on primary communicative genes PPI network: colors red and blue indicate module 1 and 2 (related to cell cycle and Immune system genes), respectively. (c) Top communicative genes subgraphs for each module. (d) Inter module top communicative genes subgraph.

The ranges of micro level network metrics and the number of genes for each metric are demonstrated in Table I.

Table 1. The characteristics of analysis of 1481-gene (all genes) PPI network based on micro level network metrics

| Micro level network metrics | Min value (all genes) | Max value (all genes) | Min value (extracted genes) | Max value (extracted genes) | No. of extracted genes |
|---|---|---|---|---|---|
| Degree | 1 | 182 | 50 | 182 | 113 |
| Stress | 0 | 2118086 | 200000 | 2118086 | 69 |
| Eigenvector | 0 | 0.13 | 0.05 | 0.13 | 86 |

3.4. PPI network of Primary communicative genes reconstruction and analysis

The statistics of 149-gene PPI network are described in Figure 5.

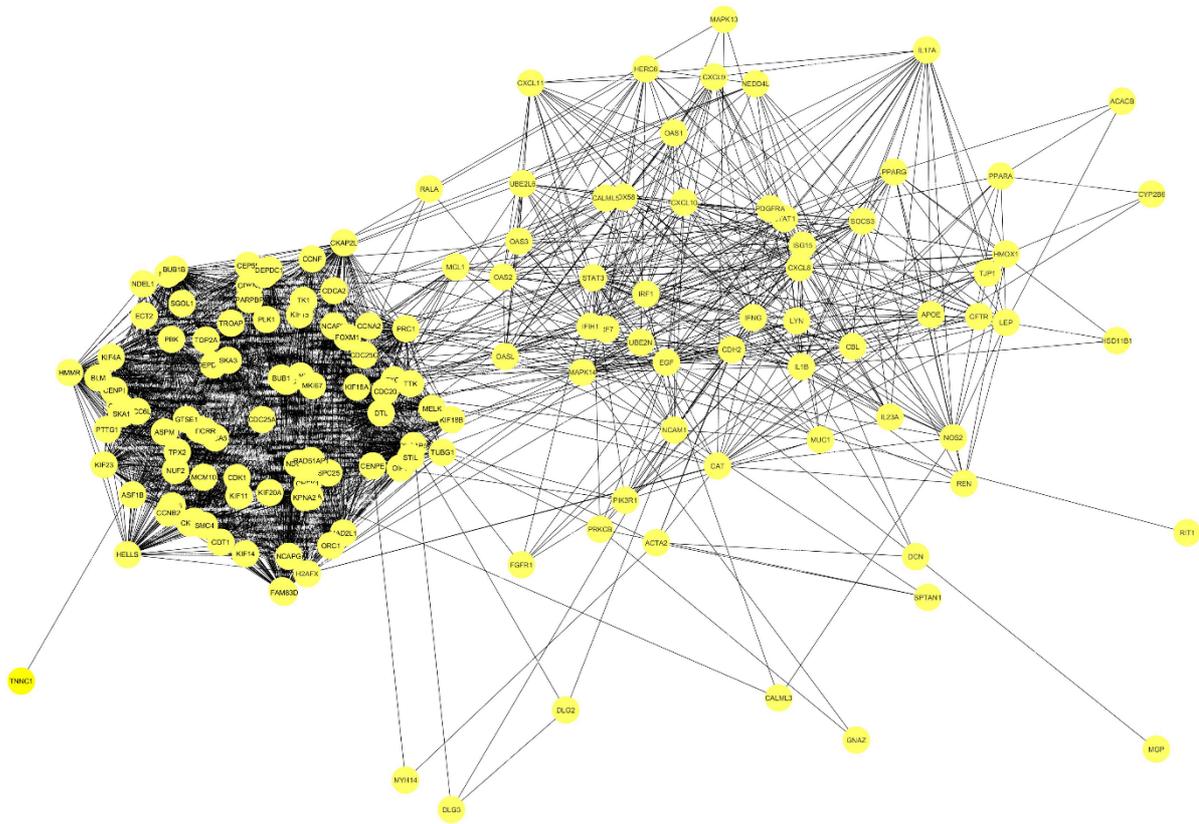

Clustering coefficient : 0.734
Connected components : 1
Network diameter : 5
Network radius : 3
Network centralization : 0.300
Shortest paths : 22052 (100%)
Characteristic path length : 1.985
Avg. number of neighbors : 48.225

Number of nodes : 149
Network density : 0326
Network heterogeneity : 0.652
Isolated nodes : 0
Number of self-loops : 0
Multi-edge node pairs : 0
Analysis time (sec) : 0.067

**Figure 5**. Primary communicative genes PPI network analyzed by Cytoscape.

3.5. Macro level network analysis

Two high density modules were recognized as a result of modularity detection on 149-connected node PPI network (Figure 4b). Color red and blue indicate module 1 including almost 58% and module 2 with almost 42% of 149 genes, respectively. Studying the roles of the genes in each module showed that the genes comprising module 1 were closely related to cell cycle, of which 81 genes were upregulated and five genes were downregulated. On the other hand, module 2 consisted of 63 genes appearing to be most likely relevant to immune system, of which 43 genes were upregulated and 20 genes were downregulated.

3.6. Top communicative genes selection

Figure S4 depicts the ranges of all metrics in each module.

3.6.1. Module 1

Table 2 presents the summary statistics for significant genes selected by filtering micro level metrics in module 1. Top communicative genes in this module were recognized with minimum degree 87, moreover, a minimum threshold of 8732 and 0.03 were employed for stress and betweenness metrics, respectively. These cut-offs were selected according to prior biological evidence on identified genes (references available in discussion) and statistical information mentioned in methods section (macro level network analysis followed by top communicative genes identification). Ranked values of obtained genes were set out in Table 3. The genes shared among all three groups including CCNA2, CCNB1 (the highest-ranked value of all gene groups), MKI67, CDK1, CDC20 and FOXM1 as top communicative genes have been shown in subgraph 1 (Figure 4c).

Table 2. The characteristics of analysis on 86-gene module (module 1) based on micro level network metrics.

| Metrics | Min | Max | Min (top) | Max (top) | No. of Total | Genes (descending order) |
|---|---|---|---|---|---|---|
| Degree | 1 | 92 | 87 | 92 | 4 | CCNB1, CDC20, CDK1, CCNA2 |
| Stress | 0 | 15636 | 8732 | 15636 | 6 | CCNB1, MKI67, CDK1, CCNA2, FOXM1, CDC20 |
| Betweenness | 0 | 0.05 | 0.03 | 0.05 | 4 | CCNB1, CDC20, MKI67, CDK1 |
| Shared genes of all groups | | | | | 5 | CCNA2, CCNB1, MKI67, CDK1, CDC20 |

Table 3. Top communicative genes (module 1) ranked based on a. degree, b. Betweenness, c. Stress

| | a. | | b. | | c. | |
|---|---|---|---|---|---|---|
| Rank | Gene symbol | Degree | Gene symbol | Betweenness | Gene symbol | Stress |
| 1 | CCNB1 | 92 | CCNB1 | 0.05 | CCNB1 | 15636 |
| 2 | CDC20 | 91 | CDC20 | 0.04 | MKI67 | 13412 |
| 3 | CDK1 | 90 | MKI67 | 0.04 | CDK1 | 10020 |
| 4 | CCNA2 | 87 | FOXM1 | 0.03 | CCNA2 | 9618 |
| 5 | | | | | FOXM1 | 8920 |
| 6 | | | | | CDC20 | 8732 |

The Main biological functions of top communicative genes in module 1 are demonstrated in Table S4.

- Enriched pathway Analysis (module 1)

Among enriched pathways associated with module 1 (based on combined score) significant pathways through KEGG and WikiPathways 2019 HUMAN databases are listed in Table S5 and Table S6.

Table 4. The characteristics of analysis on 63-gene module (module 2) based on micro level network metrics.

| Metrics | Min | Max | Ave | Min (top) | Max (top) | Total | Genes (descending order) |
|---|---|---|---|---|---|---|---|
| Degree | 1 | 43 | 15.07 | 26 | 43 | 9 | H2AFX, STAT3, EGF, IL1B, STAT1, IFNG, CXCL8, MAPK14, CXCL10 |
| Stress | 0 | 29876 | 2.67e+03 | 8518 | 29876 | 4 | STAT3, EGF, MAPK14, IL1B |
| Eigenvector | 0.000003 | 0.05 | 0.003 | 0.007 | 0.05 | 6 | H2AFX, MCL1, STAT3, EGF, UBE2N, MAPK14 |
| Betweenness | 0 | 0.07 | 0.007 | 0.04 | 0.07 | 3 | EGF, STAT3, MAPK14 |
| Shared genes of all groups | | | | | | 2 | EGF, STAT3 |

3.6.2. Module 2

All cut-off criteria of degree ≥26 (almost top 15% of high degree values), eigenvector ≥ 0.007, stress ≥ 8518 and betweenness≥0.04 were considered as the thresholds for significance to extract top communicative genes in module 2 among 63 genes. The subgraph of relationships between 11 identified top communicative genes is apparent in Figure 4c. Table 4 compares summary statistics for module 2 analysis in which we can see that overlapping genes between all groups are EGF and STAT3. Ranked values of these genes for each metric are presented separately in Table 5.

Table 5. Top communicative genes (module 2) ranked based on a. degree, b. eigenvector, c. stress, d. betweenness

| | a. | | b. | | c. | | d. | |
|---|---|---|---|---|---|---|---|---|
| Rank | Gene symbol | Degree | Gene symbol | Eigenvector | Gene symbol | Stress | Gene symbol | Betweenness |
| 1 | STAT3 | 43 | H2AFX | 0.05 | STAT3 | 29876 | EGF | 0.07913 |
| 2 | H2AFX | 43 | MCL1 | 0.018 | EGF | 27610 | STAT3 | 0.06636 |
| 3 | EGF | 40 | STAT3 | 0.015 | MAPK14 | 10910 | MAPK14 | 0.044914 |
| 4 | IL1B | 36 | EGF | 0.014 | IL1B | 8518 | | |
| 5 | STAT1 | 34 | UBE2N | 0.012 | | | | |
| 6 | IFNG | 31 | MAPK14 | 0.007 | | | | |
| 7 | 'CXCL8' | 31 | | | | | | |

Main biological functions are revealed for top 11 communicative genes in module 2 in Table S7.

- Enriched pathway analysis (module 2)

Corresponding pathway enrichment results were obtained from WikiPathways and Kegg 2019 HUMAN databases (Table S8, S9). As can be compared, overlapping pathways between KEGG and WikiPathways databases consist of RIG-I-like receptor signaling pathway, Toll-like receptor signaling pathway, AGE-RAGE signaling pathway in diabetic complications and TGF-beta signaling pathway.

3.7. Identification of pathways associated with genes in both modules 1 and 2

Pathway enrichment on combined genes from modules 1 and 2 were performed by KEGG and WikiPathways databases. Subsequently, top 20 pathways were obtained with the highest combined score. Those pathways containing intersectional genes among modules 1 and 2 were chosen (Table 6 and Table 7).

**Table 6.** Intersectional pathways among top 20 enriched pathways based on 149 merged genes from modules 1 and 2 sorted by combined score by KEGG 2019 HUMAN Database

| Index | Term | Genes | Combined score |
|---|---|---|---|
| 1 | Progesterone-mediated oocyte maturation | PLK1;PIK3R1;CDC25C;MAPK14;CDC25A;AURKA;MAPK13;CCNA2;CCNB2;CCNB1;CDK1;BUB1;MAD2L1 | 500.8 |
| 2 | Influenza A | CXCL8;PRKCB;DDX58;STAT1;PIK3R1;MAPK14;MAPK13;IFIH1;SOCS3;CXCL10;IFNG;OAS1;OAS2;IL1B;OAS3;IRF7;KPNA2 | 428.6 |
| 3 | Oocyte meiosis | CDC20;CCNB2;CCNB1;CALML5;PTTG1;PLK1;CDK1;CALML3;CDC25C;BUB1;AURKA;MAD2L1 | 290.6 |
| 4 | Cellular senescence | CXCL8;CALML5;CALML3;PIK3R1;FOXM1;MAPK14;CDC25A;MAPK13;CCNA2;CCNB2;CCNB1;CHEK1;CDK1 | 242.5 |
| 5 | Epstein-Barr virus infection | LYN;DDX58;STAT1;STAT3;ISG15;PIK3R1;MAPK14;MAPK13;CCNA2;CXCL10;OAS1;OAS2;OAS3;IRF7 | 203.9 |
| 6 | Hepatitis B | IFIH1;CCNA2;CXCL8;PRKCB;DDX58;STAT1;STAT3;IRF7;BIRC5;PIK3R1;MAPK14;MAPK13 | 192.5 |
| 7 | FoxO signaling pathway | CCNB2;CCNB1;EGF;PLK1;CAT;STAT3;PIK3R1;MAPK14;MAPK13 | 130.4 |

**Table 7.** Intersectional pathways among top 20 enriched pathways based on 149 merged genes from modules 1 and 2 sorted by combined score by WikiPathways 2019 HUMAN Database

| Index | Term | Genes | Combined score |
|---|---|---|---|
| 1 | IL-4 Signaling Pathway WP395 | SOCS3;STAT1;STAT3;BIRC5;PIK3R1;CBL;MAPK14 | 273.8 |
| 2 | Integrated Cancer Pathway WP1971 | BLM;STAT1;CHEK1;PLK1;CDK1;CDC25A | 255.5 |
| 3 | Hepatitis C and Hepatocellular Carcinoma WP3646 | CXCL8;RRM2;NOS2;STAT3;BIRC5;MAPK14 | 218.7 |
| 4 | Photodynamic therapy-induced AP-1 survival signaling. WP3611 | CCNA2;PDGFRA;IFNG;MAPK14;MCL1;MAPK13 | 212.4 |
| 5 | Regulation of toll-like receptor signaling pathway WP1449 | CXCL10;CXCL11;CXCL9;CXCL8;STAT1;IL1B;PLK1;IRF7;PIK3R1;MAPK14;MAPK13 | 198.7 |

3.8. Inter top communicative genes subnetwork interactions

The subnetwork including the top communicative genes, from the two modules, is summarized in Figure 4d that may act as molecular signatures for underlying phenotypes of psoriasis disease. The overall numbers of 6 and 11 genes (17 in total) for modules 1 and 2, respectively, were identified among a total number of 149 genes as top communicative.

## 4. Discussion

Through the reconstruction of a PPI network using DEGs followed by network analysis, 152 primary communicative genes were chosen from the initial 1481 connected nodes in PPI. As the final step of network analysis, on the basis of the macro-level metrics, primary communicative genes PPI network was shown to belong in two distinct modules comprising cell cycle (module 1) or immune system related genes (module 2). Each module was examined with different network micro-level metrics in order to find the top communicative genes. Module 1 top communicative genes contained CCNA2, CCNB1, CDC20, CDK1, FOXM1 and MKI67; and module 2 involved STAT3, EGF, H2AFX, IL1B, IFNG, STAT1, CXCL8, CXCL10, MAPK14, MCL1 and UBE2N (Table 3, 5).

Below, the top communicative genes for which there are experimental evidence relevant to psoriasis are discussed, while mentioning those psoriasis-related intra- and inter-module interactions for which there is experimental evidence (Figure 4c,d).

4.1. Module 1 (cell cycle)

CCNB1 had the highest betweenness, degree and stress measures (Table 3). It was demonstrated by J.L. Melero et al. to have a higher expression in psoriatic tissue together with three other genes i.e. CCNA2, CCNE2 and CDK1, which could lead to uncontrolled cell proliferation [8].

CDK1 expression is increased in human psoriatic lesions [26]. It can bind to CCNB1 [27], CCNA2 [28] which have been categorized in module 1 top communicative genes and CCNB2 [29], which exists in module 1.

CCNA2 along with two other cyclins i.e. CCNB1 and CCNB2 involved in cell proliferation in addition to cell division cycle genes, *CDC2* and *CDC20,* are up-regulated in psoriatic skin [30].

FOXM1 is one of the core transcription factor regulators in psoriasis [31]. It is the only down regulated top communicative gene.

CDC20 is an activator of anaphase-promoting complex or cyclosome (APC/C) and plays a critical role in mitosis and S phase through the degradation of S phase and mitosis cyclins, which leads to the exit from mitosis. APC/C phosphorylation by CDK1-cyclin B1 leads to the induced binding of CDC20 to APC/C. The activated APC/C$^{cdc20}$ target, cyclin B1 APC/C, also promotes the start of chromosome segregation in metaphase-anaphase transition by degrading a protein that inhibits anaphase [32].

4.2. Module 2 (Immune system)

EGF exerts its effects through binding to an EGF/TGF alpha receptor in responsive cells. Any alteration in EGF binding pattern has shown to lead to abnormal differentiation and growth found in diseases such as psoriasis [33]. Apoptosis inhibition and keratinocyte hyperproliferation are evident in psoriasis. Several proteins are present in these mechanisms and one of them is EGF [34]. EGF is upregulated in lesional psoriatic skin in comparison to non-lesional in our results. Since keratinocytes bear receptors for the majority of psoriasis-signature cytokines, they represent the "key responding" tissue cells to the psoriatic microenvironment. They respond to psoriatic cytokines by proliferating and amplifying inflammation through the production of other cytokines (i.e., IL-1F9, TNFα, IL-17C, IL-19, TSLP), chemokines proliferation-stimulating factors, and other pro-inflammatory products, such as AMPs [35].

STAT3, p-STAT3, UBE2N and CDK6 are downregulated by the overexpression of hsa-miR-4516, which are consistently upregulated in psoriasis and induce apoptosis in HaCaT cells [36].

According to our analysis, STAT3 is upregulated, which explains more severe conditions in lesional relative to non-lesional conditions.

IL1B is a key factor in cutaneous inflammation and plays an axial role in the instigation of an inflammatory Th17 micro-milieu in psoriasis and other autoinflammatory diseases. Its biosynthesis is controlled at transcription level followed by ensuing posttranslational process by means of inflammatory caspases. Zwicker et al. detected that inflammatory caspase-5 is active in psoriatic skin lesions and epidermal keratinocytes. Besides, they showed that IFNG and IL17A cooperatively triggered caspase-5 expression in keratinocytes culture and it was dependent on psoriasis, which is an antimicrobial peptide (S100A7) [37].

IFNG, i.e. single type II IFN, plays a key role in the trigger and escalation of systemic autoimmunity. In conjunction, the interferons affect a wide range of biological responses such as anti-tumor effects, yielding protection against bacterial and viral infections and regulation of effector cells in adaptive and innate immune responses [38]. Using microarray analysis of *in-vitro* derived macrophages which were treated with IFNG, Fuentes-Duculan et al. showed that a plethora of the genes upregulated in macrophages were present in psoriasis, namely, HLA-DR, STAT1, CXCL9 and Mx1 [39]. The importance of IFNG in autoimmune responses is in compliance with its stunningly high overexpression observed in lesional compared to non-lesional samples.

CXCL8 is secreted by endothelial cells, monocytes, fibroblasts, macrophages and lymphocytes. This chemokine acts as a bridge between the cell cycle and immunity leading to the activation of keratinocytes of the lesional skin to produce inflammatory cytokines. It can also participate in the angiogenesis and migration of neutrophils and T cells in the inflammation site. CXCL8 is regulated by other cytokines such as IL17A and IL1B, which is critical for the pathogenesis of psoriasis and influences lesional keratinocytes of psoriasis [39, 41].

IL12B codes the p40 subunit of IL-23 and IL-12, cytokines playing key roles in Th17 and Th1 procedures, respectively. The alterations in this gene significantly enhances the risk of psoriasis. Johnston et al. demonstrated that a psoriatic risk haplotype at IL12B locus triggers an increase in the expression of IL12B in monocytes. This increase is correlated with elevated serum levels of IFNG , IL-12 and CXCL10, which is an IFNG induced chemokine [43].

UBE2N has roles in cellular processes by contributing to the immune system, DNA replication and repair. UBE2N accounts for the ubiquitination of TRAF family which regulates Toll-like receptors on dendritic cells, macrophages and neutrophils in NFkB pathway of the innate immune system. Therefore, UBE2N could contribute to the inflammatory responses [42, 43].

H2AFX expression is shown to be elevated in psoriatic arthritis patients compared to control ones through proteomic analysis of synovial fluid. Therefore, H2AFX is considered as a potential biomarker ensuing serum investigations [46].

STAT1 expression and activity are both considerably increased in lesional psoriatic skin conditions compared to non-lesional psoriatic skin [47].

MCL1 encodes an anti-apoptotic protein, myeloid cell leukemia-1. Ultraviolet B radiation therapy inhibits growth and induces apoptosis of human skin cells by downregulating MCL1 expression [48]. This gene shows upregulation in our study.

MAPK14 and LYN are two instances of protein kinase and signal integrators through which, immune cascades are channelized. They are expressed in higher amounts in lesional psorisis [49].

4.3. Pathway enrichment

In order to achieve a deeper mechanistic insight into the interactions of primary communicative genes, we performed pathway enrichment analysis. Based on Wikipathways and KEGG pathways enrichment analysis on module 1, top five enriched pathways from either database were Cell cycle, Progesterone-mediated oocyte maturation, Oocyte meiosis, p53 signaling pathway, Cellular senescence, Regulation of sister chromatid separation at the metaphase-anaphase transition, Retinoblastoma Gene in Cancer, Gastric Cancer Network 1 and ATM Signaling Pathway. Likewise, for module 2 Type II interferon signaling (IFNG), IL-10 Anti-inflammatory Signaling Pathway, Influenza A, Pertussis, Hepatitis C, Toll-like Receptor Signaling Pathway and RIG-I-like Receptor

Signaling were enriched (Table 5,6,8,9). Thereafter, based on pathway enrichment analysis for 149 merged genes from module 1 and 2, intersectional pathways having the largest number of top communicative genes would be Influenza A, Cellular senescence, Epstein-Barr virus infection, Hepatitis B, Photodynamic therapy-induced AP-1 survival signaling and Regulation of toll-like receptor signaling pathway (Table 6,7).

## 5. Conclusions

In conclusion, despite the existence of numerous medications devised for soothing the symptoms of psoriasis, no ultimate treatment has been achieved for the disease. Palpably, the determination of pivotal genes that play important roles in the pathogenesis of psoriasis is critical for the development of diagnostic/therapeutic targets. Owing to the fact that out of 17 mentioned genes, all of them have been experimentally proven to be related to psoriasis, we do believe that it is a high likelihood that the other members of these modules are also related to psoriasis. Additionally, only one of the top communicative genes, i.e. CXCL8, was among top 100 differentially expressed genes (Table S2, S3), asserting that crucial genes found through our network analysis pipeline wouldn't be straightforwardly identified merely relying on DEGs or without incorporating macro- and micro-level network metrics. Thereby, primary communicative genes found in our study and not investigated elsewhere in connection to psoriasis are put forward for further demonstrations (Table S10).

Subsequent large-scale validation of the latter in serum is critical in order to prove these proteins as putative biomarkers or/and therapeutic targets for the diagnosis and/or the treatment of psoriasis. Beyond that, since there are several experimental studies in which combinatorial gene sets have been used as therapeutic/diagnostic targets [48, 49] , from a network science point of view leading to our intra- and inter-module interactions asserted by experimental studies, it is likely that the mentioned top communicative subnetworks unveil disease-specific patterns.

**Supplementary Materials:** The following are available online at www.mdpi.com/xxx/s1, Figure S1: The heatmap of the Pearson's correlation coefficient (r2) of 170 samples as pairwise comparisons for quality control. Figure S2: Boxplots of raw data for the expression level values transformed to log base 2. (a) And (b) demonstrate pre-normalized and normalized samples, respectively. Figure S3: Volcano plot representing the result of hypothesis testing between lesional and non-lesional samples. Figure S4: Metrics distribution for each module. Table S1: Top 2000 unique DEGs. Table S2: Top 50 upregulated DEGs among lesional and non-lesional samples. Table S3: Top 50 downregulated DEGs among lesional and non-lesional samples. Table S4: Main biological functions of top six communicative genes in module 1. Table S5: Top 10 enriched pathways in module 1 ranked in order of combined score by KEGG 2019 HUMAN Database. Table S6: Top 10 enriched pathways in module 1 in order of combined score by WikiPathways 2019 HUMAN database. Table S7: Main biological functions of 11 top communicative genes in module 2. Table S8: Top 10 enriched pathways in module 2 in order of combined score by Wikipathways 2019 HUMAN database. Table S9: Top enriched pathways in module 2 ranked in order of combined score by KEGG 2019 HUMAN Database. Table S10: 149 primary communicative genes.

**Conflicts of Interest:** The authors declare no conflict of interest.